# Improving Software Developer's Competence: Is the Personal Software Process Working?

Pekka Abrahamsson[1], Karlheinz Kautz[2], Heikki Sieppi[3] and Jouni Lappalainen[3]

[1] VTT Technical Research Centre of Finland, VTT Electronics,
P.O.Box 1100, FIN-90571 Oulu, Finland
`Pekka.Abrahamsson@vtt.fi`

[2] Department of Informatics, Copenhagen Business School,
Howitzvej 60, 3., DK-2000 Frederiksberg, Denmark
`Karl.Kautz@cbs.dk`

[3] Department of Information Processing Science,
P.O.Box 3000, FIN-90014 University of Oulu, Finland
`{Heikki.Sieppi; Jouni.Lappalainen}@oulu.fi`

**Abstract.** Emerging agile software development methods are people oriented development approaches to be used by the software industry. The personal software process (PSP[SM]) is an accepted method for improving the capabilities of a single software engineer. Five original hypotheses regarding the impact of the PSP to individual performance are tested. Data is obtained from 58 computer science students in three university courses on the master level, which were held in two different educational institutions in Finland and Denmark. Statistical data treatment shows that the use of PSP did not improve size and time estimation skills but that the productivity did not decrease and the resulting product quality was improved. The implications of these findings are briefly addressed.

## 1 Introduction

The emergence of agile software development technologies such as Scrum [1] and Extreme Programming [2] has brought the attention back to the individual software developer's competence. Agile software development solutions place emphasis on self-directing development teams and thus rely on the abilities of individual software engineers to make f. ex. informed choices about the development methodology they intend to use. Agile development methods, however, do not provide guidance on how to develop and maintain such a competence.

The personal software process (PSP) is an accepted (and to our knowledge the only) method for improving software processes at the personal level [3]. The PSP method is essentially about the individual software engineer's ability to learn to control and to develop his own development processes. Only after having explored different techniques an engineer is able to decide upon his most effective personal solution [4]. Furthermore, the use the PSP indicates increased personal responsibility for quality and productivity improvements. While the software engineering research is keen in introducing new and enhanced methods, often the evaluation of existing ones is limited [5].

The purpose of this paper is to test the original hypotheses made by some PSP proponents to investigate whether the basic claims are supported by the data collected from three courses attended by computer science students on the master level and held in two different institutions representing different cultural settings and contexts[1]. We have reported the experiences from the first two of the courses elsewhere [6, 7]. Considering the validity of using students as research subjects, it has been shown that they are valid representatives for practitioners in industry [8].

The baseline for the hypotheses was Hayes and Over's [9] PSP impact study where they obtained data from 298 engineers. Wesslén [10] replicated Hayes and Over's study and confirmed their results. Our results support Hayes and Over's findings in terms of increased product quality and in terms of witnessing no change in the productivity levels. Our results offer contrasting results regarding the improvement in size and effort estimation skills.

The paper is organised as follows. The following section provides a brief overview of the PSP method. This is followed by an introduction to the research settings. The results from the statistical tests are presented in section 4 and briefly addressed in the concluding section.

## 2. Overview of the PSP

The PSP was developed by Humphrey [4] to extend the improvement process from the organisational level to the individual software engineer. The aim of the PSP is to enable software engineers to control and manage their software processes and products as well as to improve their predictability and quality. This is achieved through the gradual introduction of new elements into the baseline personal process. The PSP is usually taught in form of an educational course with a number of programming assignments. A student entering a PSP course starts with PSP0, that is, his current process enhanced with time and defect tracking instruments. The subsequent PSP levels extend the personal baseline process to include a coding standard, software size measurement and test reporting practices to name a few. The size and effort estimations are performed using a PSP defined estimation method, where the students systematically use the historical data they have collected from their programming exercises during the course. At PSP2 level the focus is directed towards personal quality management through the introduction of code and design review practices. Students develop their personal defect and design review checklists, based on their historical defect data. Finally, PSP3 scales up the process from a single module development to larger scale projects. As an outcome, the project is divided in a series of smaller sub projects that are then incrementally implemented.

---

[1] These differences are not addressed in this paper however. Instead, we use the data material in total.

## 3. Research Settings

The PSP data used in this paper is collected from three PSP courses held in University of Oulu in (fall 2000; Oulu1, spring 2002; Oulu2) and at Copenhagen Business School in fall 2001; CBS1. The details of the research settings are highlighted in Table 1.

With regard to the original proposal for conducting PSP courses, the courses were adjusted and redesigned based on experiences drawn from the earlier courses. The following changes were implemented: The assignment contexts were tailored to fit the local environment. The second major change was in the target PSP level to be achieved. In the first course, the target was to reach the PSP2 level. In the second course and third course the goal was to reach the PSP3 level. The principal change from CBS1 to Oulu2 was the redesign of the course assignments; as a consequence the size of the development work was doubled on average.

Table 1. Details of the research settings

|  | Oulu1 | CBS1 | Oulu2 |
|---|---|---|---|
| Lectures | 9 | 13 | 10 |
| Course length (weeks) | 10 | 14 | 10 |
| Programming assignments | 7 | 8 | 8 |
| Reports | 3 | 2 | 2 |
| PSP level achieved | 2 | 3 | 3 |
| Total number of participants | 31 | 22 | 32 |
| Pass % | 65% | 77% | 53% |
| Study year | 3-4 | 4-5 | 3-5 |
| Time used for a single programming assignment | 5.23 (median) 23.85 (max) | 5.08h (median) 26.27 (max) | 6.35h (median) 59.33 (max) |
| Work size in lines of code for a single programming assignment | 98.50 (median) 571 (max) | 101.25 (median) 855 (max) | 209 (median) 1741 (max) |
| Main implementation languages used | Java, C++ | Java, C++, Visual basic | Java |

For each assignment, the students had a full week to complete the work and submit the results. Johnson [11] has found that the data collected from a PSP course is often error prone. Thus, in order to ensure the validity of the data collected each assignment was rigorously checked and feedback provided. All data inconsistencies were reported and clarified with the student through email communication. The data collection process was facilitated through the use of electronic documents. Automated data collection tools were used in the Oulu2 case. Time and defect tracking was then automated with suitable self-administrated tools.

## 4. Results

In this section the data administration procedures, hypotheses tested and the results from the statistical tests are presented.

### 4.1. Data administration

The data obtained from the three courses were organised and all questionable data points were removed from the data samples. Box plot diagrams that show the 5 number summary of a data set (median, min, max, upper quartile and lower quartile values) were drawn for visualisation purposes. Outliers were subsequently removed (max 5% of samples) from each of the data set in accordance to statistical rules [12]. All hypotheses are were tested using standard t-test assuming unequal variances or a paired t-test as described in brief in [13]. T-test is a parametric test that can be used to compare two independent samples. In our case, the independent data samples for each hypothesis were the PSP0 level and PSP2/3 levels.

### 4.2. Results from the statistical analysis

The visualisation of the results (Figure 1) leads to suspect that while no improvement can be identified in size or time estimation skills, the productivity appears to stay roughly at the same level, overall defect density is lower and the percentage of defects removed before compile is significantly higher.

The hypotheses were drawn from a PSP impact study administrated by Hayes and Over [9]. They base their claims on PSP data collected from 298 engineers. Table 2 contrasts Hayes and Over's findings against the data obtained from 58 students in this study.

Table 2. The results

| Hayes and Over findings | This study |
|---|---|
| A median of 2.5x improvement in size estimation | Not supported: No change or improvement |
| A median of 1.75x improvement in effort estimation | Not supported: No change or improvement |
| A median of 1.5x reduction in total defect density | Supported: 1.85x reduction in total defect density |
| A median of 50% increase in percentage of defects removed before compile | Not supported: 3.5x improvement but due to high variance insignificant |
| No statistically significant change in productivity | Strongly supported |

Statistical tests, however, show support only for the productivity (statistically strong support) hypotheses and overall defect density reduction (1.85x) when tested at 0.05 level. The variance is too high for testing the percentage of defects removed before compile. No statistical support can be given even though visually the difference is clearly noticeable.

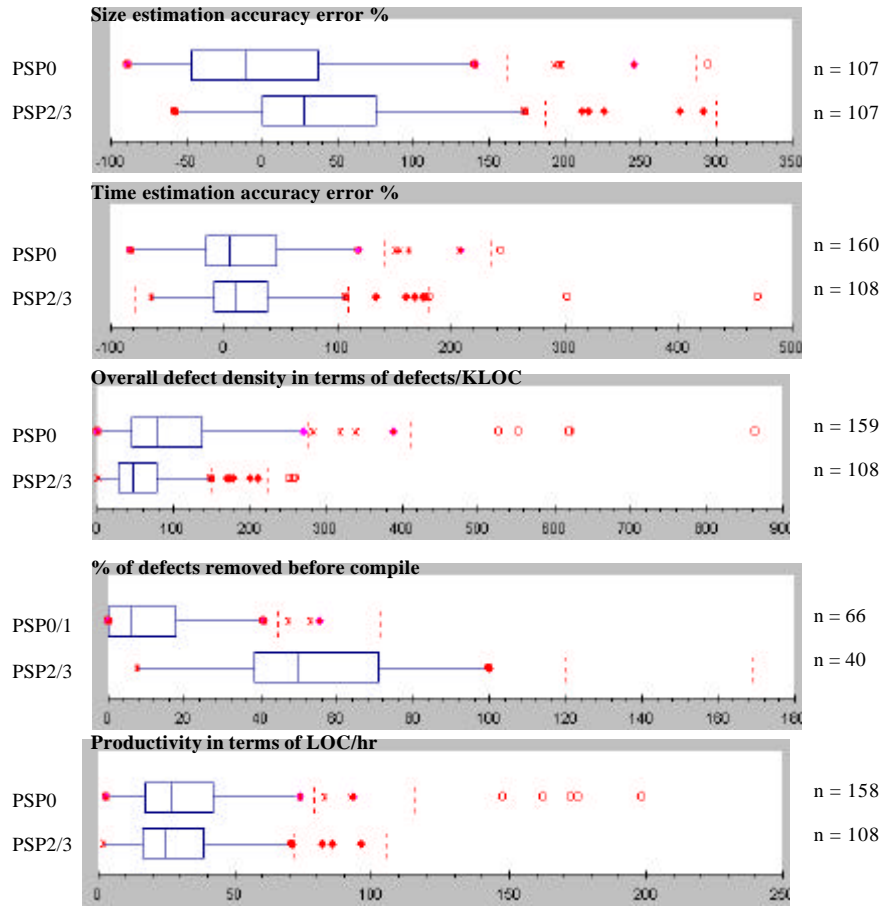

**Figure 1. Box plot diagrams of the results**

## 5. Conclusions

The PSP is the only method targeted at individual software engineers' competence development. This study tested the original claims made by PSP proponents. The data was obtained from a total of 58 students. The results show that PSP does not enable immediate improvement in size or time estimation skills but that the product quality tends to be higher through the use of the method. A significant finding remains that while the PSP is claimed to be a heavy and documentation oriented process it does not reduce the productivity, which bears significance when considering the use of the method or a similar approach. It is our suggestion that PSP data collection abilities should be incorporated to the agile software development methods, which rely on the abilities of a single software engineer. A tailored version of PSP would provide the essential techniques and structure to systematically pursue improvements. In conclusion, several new software development methods are introduced on a yearly basis. However, while software professionals seek a rational basis for making a decision which method they should adopt, the basis for such a rationalization is missing. Methods introduced continue be based more on faith than on an empirical data [14]. There is no quick solution to the problem described. Fenton [14] suggested that only by contributing gradually to the empirical body of knowledge within the specific area of application we as researchers are able to test the basic software engineering hypotheses made. Our principal aim in this paper, therefore, has been to contribute to the empirical body of knowledge within the area of software engineering and in specific within the area of personal competence development. It is our firm belief that researchers and practitioners are better equipped to rationally evaluate f. ex. the use of the PSP on the basis of data as reported in this paper.

## References

[1] K. Schwaber and M. Beedle, *Agile Software Development With Scrum.* Upper Saddle River, NJ: Prentice-Hall, 2002.

[2] K. Beck, *Extreme programming explained.* Reading, MA.: Addison Wesley Longman, Inc., 2000.

[3] S. Zahran, *Software process improvement: practical guidelines for business success.* Reading, Mass.: Addison-Wesley Pub. Co., 1998.

[4] W. S. Humphrey, *A discipline for software engineering.* Reading, Mass.: Addison Wesley, 1995.

[5] K. E. Wiegers, "Read my lips: No new models," *IEEE Software*, vol. 15, pp. 10-13, 1998.

[6] P. Abrahamsson and K. Kautz, "Personal Software Process: Classroom experiences from Finland," presented at 7th European Conference on Software Quality, Helsinki, Finland, 2002.

[7] P. Abrahamsson and K. Kautz, "Personal software process: Experiences from Denmark," presented at Euromicro 2002, Dortmund, Germany, 2002.

[8] M. Höst, B. Regnell, and C. Wohlin, "Using students as subjects - a comparative study of student and professionals in lead-time impact assessment," *Empirical Software Engineering*, vol. 5, pp. 201-214, 2000.

[9] W. Hayes and J. W. Over, "The Personal Software Process (PSP): An Empirical Study of the Impact of PSP on Individual Engineers," Software Engineering Institute, CMU/SEI-97-TR-001, Technical Report CMU/SEI-97-TR-001, http://www.sei.cmu.edu/publications/documents/97.reports/97tr001/97tr001abstract.html, 1997.

[10] A. Wesslén, "A replicated empirical study of the impact of the methods in the PSP on indivudal engineers," *Empirical Software Engineering*, vol. 5, pp. 93-123, 2000.

Abrahamsson, P., Kautz, K., Sieppi, H., & Lappalainen, J. (2002) Improving software developer's competence: Is the Personal Software Process Working? Presented at Workshop on Empirical Software Engineering, 9.12.2002, Rovaniemi, Finland

[11]  P. M. Johnson, "The personal software process: A cautionary case study," *IEEE Software*, vol. 15, pp. 85-88, 1998.

[12]  D. C. Montgomery, *Design and analysis of experiments*, 4th edition ed. New York: John Wiley & Sons, 1997.

[13]  C. Wohlin, P. Runeson, M. Höst, M. C. Ohlsson, B. Regnell*, et al.*, *Experimentation in software engineering*. Boston: Kluwer Academic Publishers, 2000.

[14]  N. Fenton, "Viewpoint Article: Conducting and presenting empirical software engineering," *Empirical Software Engineering*, vol. 6, pp. 195-200, 2001.